\documentclass[fleqn,twoside]{article}
\usepackage{espcrc2}

\usepackage{epsfig}

\title{Using SKA to observe relativistic jets from X-ray
  binary systems}

\author{R. Fender\address[UvA]{
Astronomical Institute 'Anton Pannekoek', University of Amsterdam}}
\begin{document}

\begin{abstract}
I briefly outline our current observational understanding of the
relativistic jets observed from X-ray binary systems, and how their
study may shed light on analogous phenomena in Active Galactic Nuclei
and Gamma Ray Bursts. How SKA may impact on this field is sketched,
including the routine tracking of relativistic ejections to large
distances from the binaries, detecting and monitoring the radio
counterparts to 'quiescent' black holes, and detecting the radio
counterparts of the brightest X-ray binaries throughout the Local
Group of galaxies.
\end{abstract}
\vspace{1pc}

\maketitle

\section{Introduction}

\subsection{Jets from X-ray binaries}

X-ray binary systems (Lewin \& van der Klis 2004; Lewin, van Paradijs
\& van den Heuvel 1995) are the sites of the most dramatic, ongoing,
high-energy astrophysical phenomena on non-cosmlogical
scales. Compact, relativistic stars -- neutron stars or 'stellar mass'
black holes -- accrete material from a binary companion in a
relatively short (hours to weeks) period double system. An earlier
brief discussion on the potential of SKA for the study of these
objects was presented in Fender (1999).

The key signature of the relativistic accretion process in the radio
band are the {\em jets}. As in Active Galactic Nuclei (AGN), and
probably Gamma-Ray Bursts (GRBs) these jets seem to correspond to
highly relativistic outflows of matter, probably at least in part
baryonic, from very close to the accreting central object (in some
cases the launch point may be as close as a handful of gravitational
radii, $G M / c^2$). As with AGN and GRBs the emission mechanism is
almost certainly synchrotron from relativistic electrons spiralling in
magnetic fields. These synchrotron-emitting clouds of electrons have
themselves relativistic bulk motions, along a (more or less) fixed
axis, and this is the jet. Fig 1 provides a sketch of the probable
geometry of a jet-producing X-ray binary.

\begin{figure}
\psfig{figure=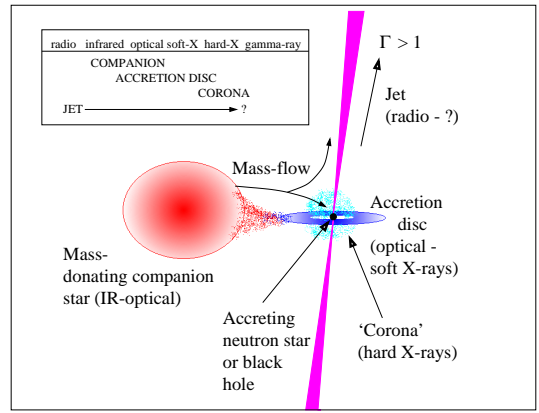,angle=270,width=7cm}
\caption{A schematic diagram of a (low-mass) X-ray binary system,
  sketching the various physical components and regions of the
  electromagnetic spectrum in which they emit. The jet is unique in
  having an extremely broad spectral component, from being the only
  major contributor to the radio emission to one of probably several
  emitting components in the X-ray band. When we detect an X-ray
  binary in the radio band we are almost certainly detecting the jet
  (see further arguments in Fender 2004).
}
\end{figure}

We know of currently 200-300 X-ray binaries (Liu, van Paradijs \& van
den Heuvel 2000,2001), probably corresponding to an underlying
population of some tens of thousands of compact objects in binary
systems in our galaxy.  The distribution of these X-ray binaries
within our galaxy has been addressed most recently by Grimm, Gilfanov
\& Sunyaev (2001) and Jonker \& Nelemans (2004). We see the brightest
X-ray binaries essentially to the other side of the galaxy. The
distribution of systems from Grimm et al. is presented in Fig
2. High-mass and low-mass X-ray binaries (where the prefix applies to
the mass of the companion star, not the accreting object) are
distributed differently, with the generally younger high-mass systems
concentrated near regions of star formation, i.e. the spiral
arms. Note however, that there does not appear to be a strong effect
on the disc-jet coupling whatever the mass of the companion star. This
is natural since the inner regions of a bright accretion disc do not
feel the influence of the companion star in any strong
way. Nevertheless, in some cases the photon field of a bright
companion star {\em might} act as a source of photons for external
comptonisation of the extended jet structure (e.g. Georganopoulos,
Aharonian \& Kirk 2002).

Over the past decade, key observations with ATCA, MERLIN, The Ryle
Telescope, VL(B)A, EVN and WSRT have provided remarkable insights into
the variable process of jet formation in these systems. When coupled
with simultaneous X-ray measurements these observations have probed
the dynamical coupling between accretion and outflow around accreting
relativistic objects in a way which is not possible for AGN (which
vary too slowly, in general) or GRBs (in which, in a sense, its all
over in an instant).

Based upon studies such as these, clear patterns have emerged in the
coupling between accretion and outflow in these sources, which we
shall outline below.

\subsubsection{Black hole X-ray binaries}

Fig 3 presents the radio vs. X-ray plane for nearly all the radio-loud
black hole X-ray binaries (a notable exception is SS 433, which evades
easy classification, but is essentially as radio-loud as Cyg X-3 at a
significantly lower apparent X-ray luminosity), scaled to a distance
of 1 kpc in order to compare luminosities (from Gallo, Fender \&
Pooley 2003). Examples are also shown of resolved jets from different
regions of the plane. Note that scaling by the estimated two orders of
magnitude sensitivity improvement expected from SKA, much more of the
plane will be accessible to direct imaging of the jets.

The 'hard' (or 'low/hard') X-ray spectral state is ubiquitous at X-ray
luminosites below about 1\% of the Eddington luminosity (typically
considering a $\sim 10M_{\odot}$ black hole), and in some sources may
persist to as bright as 10\% of this in the rising phase of an
outburst. The state gets its name from its hard X-ray spectra which
shows only a weak blackbody component which may be ascribed to an
accretion disc, but is instead dominated by a component which peaks in
most sources around 100 keV and is generally interpreted as arising in
thermal Comptonisation (see e.g. McClintock \& Remillard 2004 and
references therein for this interpretation, and Markoff, Falcke \&
Fender 2001 for a 'synchrotron-based alternative'). In this state
there is a `compact' self-absorbed jet which manifests itself as a
`flat' (spectral index $\alpha \sim 0$ where $\alpha = \Delta
\log{S_{\nu}} / \Delta \log{\nu}$) or `inverted' ($\alpha \geq 0$)
spectral component in the radio, millimetre and (probably) infrared
bands (e.g. Fender 2001, Corbel \& Fender 2002). The radio luminosity
of these jets shows a strong, non-linear correlation with X-ray
luminosity (Corbel et al. 2003; Gallo, Fender \& Pooley 2003) -- see
Fig 3 up to a scaled X-ray flux of about 1 Crab. The correlation takes
the form

\[
L_{\rm radio} \propto L_{\rm X}^b
\]

\noindent where $b \sim 0.7$. This is consistent with the 'Fundamental
Plane' (see below, and Fig 6) for sources of all approximately the
same mass.  The steady jets which we infer to exist have only been
directly spatially resolved in the case of Cyg X-1 (Stirling et
al. 2001), although the `plateau' jet of GRS 1915+105 is
phenomenologically similar and has also been resolved (Dhawan et
al. 2000; Fuchs et al. 2003). The suggestion that such steady, compact
jets are produced even at very low accretion rates (Gallo, Fender \&
Pooley 2003; Fender, Gallo \& Jonker 2003) has recently received
support in the flat radio spectrum observed from the `quiescent'
transient V404 Cyg at an average X-ray luminosity $L_{\rm X} \sim
10^{-6} L_{\rm Edd}$ (Gallo, Fender \& Hynes 2004).

Brighter sources may enter the 'soft' (or 'high/soft' or 'thermal
dominated') X-ray state at higher luminosities (typically around
10--50\% Eddington). In this state the X-ray spectrum is dominated by
a $\sim$blackbody component probably corresponding to an optically
thick and geometrically thin accretion disc extending to the innermost
stable circular orbit (McClintock \& Remillard 2004 review the
properties in more detail).  In this state the radio emission, and
probably therefore jet production, is strongly suppressed (Tanabaum et
al. 1972; Fender et al. 1999b; Gallo, Fender \& Pooley 2003). This
effect in which softer X-ray states reduce the radio emission can be
seen clearly in the interval 1--10 Crab in Fig 3. 

Fig 4 (from Tigelaar 2004) presents the broadband spectrum of the
black hole X-ray binary Cygnus X-1 in hard (labelled 'Low/Hard') and
soft(-er) (labelled 'High/soft or Intermediate') states. The change in
the X-ray spectrum from hard to soft states, resulting in a reduction
in emission around 100 keV ($\sim 10^{19}$ Hz) and an increase in the
temperature and luminosity of the accretion disc (around 1 keV $\equiv
10^{17}$ Hz), are clearly associated with a dramatic 'quenching' of
the radio component.

Additionally there are bright events associated with transient
outbursts and state transitions (of which more later), which are often
directly resolved into components displaying relativistic motions away
from the binary core (e.g. Mirabel \& Rodriguez 1994; Hjellming \&
Rupen 1995; Fender et al. 1999,2002b) not only in the radio but also -- at
least once -- in the X-ray band (Corbel et al. 2002). These events
typically display optically thin (synchrotron) radio spectra ($\alpha
\leq -0.5$). Both kinds of jets are clearly very powerful and coupled
to the accretion process. See Mirabel \& Rodriguez (1999) and Fender
(2004) for a more thorough review of the observational properties of
X-ray binary jets.

\begin{figure}
\psfig{figure=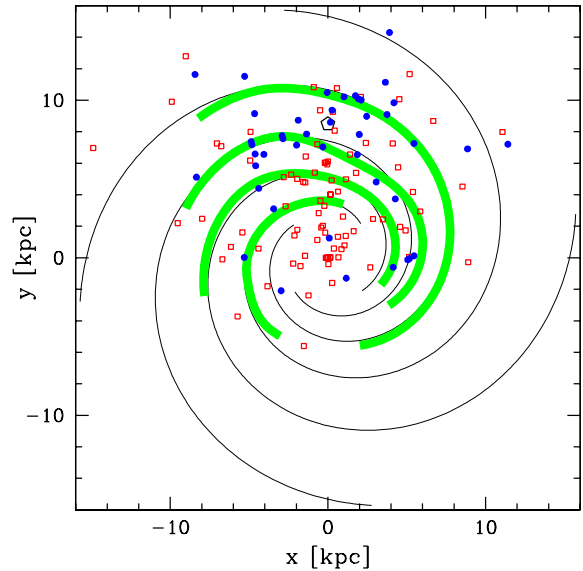,width=8cm}
\caption{The distribution of known X-ray binary systems within the
  Milky Way. From Grimm, Gilfanov \& Sunyaev (2002). Some of the
  distances may be significantly underestimated (e.g. Jonker \&
  Nelemans 2004), in which case we have probably observed sources up
  to 20+ kpc distant. The Sun is at (0,8.5) kpc. Sold circles indicate
  high-mass X-ray binaries (those with massive companions) while open
  symbols indicate low-mass X-ray binaries (typically with
  companions of a solar mass or less).
}
\end{figure}
\begin{figure*}
\centerline{\psfig{figure=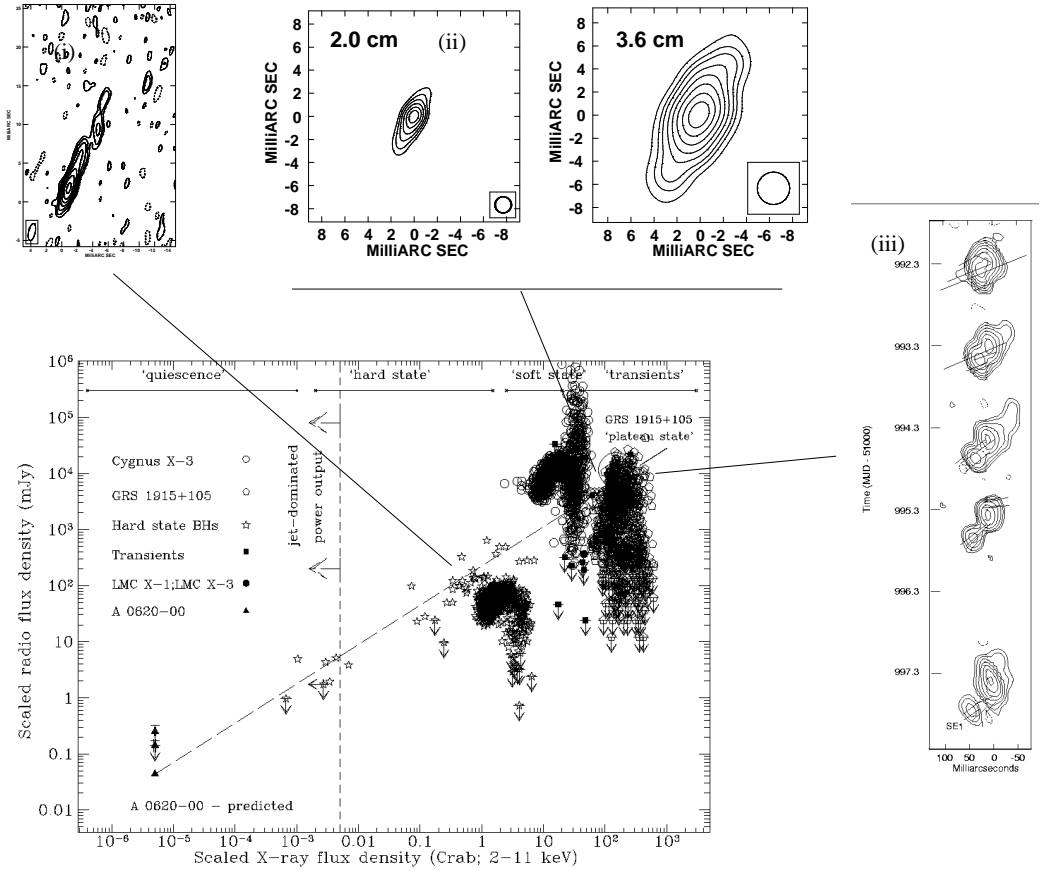,width=14cm}}
\caption{The radio : X-ray plane for galactic black hole X-ray binary
systems, with all sources scaled to a distance of 1 kpc. The brightest
transient sources have peak radio flux densities in excess of 1 Jy at
GHz frequencies. The faintest sources currently detected at about the
$\sim 0.1$ mJy level. SKA will allow us to detect sources confidently
down to the $\mu$Jy level. From Gallo, Fender \& Pooley (2003). The
insets, clockwise from top left correspond to (i) the steady jet in the
'low/hard' state from Cyg X-1 (Stirling et al. 2001) (ii) the very
powerful steady jet from GRS 1915+105 in the `plateau' hard X-ray
state (Fuchs et al. 2003) and (iii) relativistic ejections from GRS
1915+105 with spatially resolved linear polarisation (Fender et
al. 2002b). 
}
\label{gfp}
\end{figure*}

Fig 5 summarises our current best model for the phenomenology of the
disc--jet coupling in black hole X-ray binaries (from Fender, Belloni
\& Gallo 2004). In essence, we suggest that until the disc comes very
close to the black hole (around the time we observe it dominating the
X-ray spectrum) a $\sim$steady jet is formed. During transitions to
soft states as the disc is making its final 'collapse' inwards, the
jet velocity increases sharply. This results in a relativisitically
moving, optically thin, internal shock, followed by a suppression of
the 'core' radio emission while the source is in a soft X-ray state.
The seemingly ubiquitous presence of jets in hard X-ray states has
been taken as strong evidence for the magnetohydrodynamic (MHD)
production of jets (see e.g. Meier 2001). 

\begin{figure}
\psfig{figure=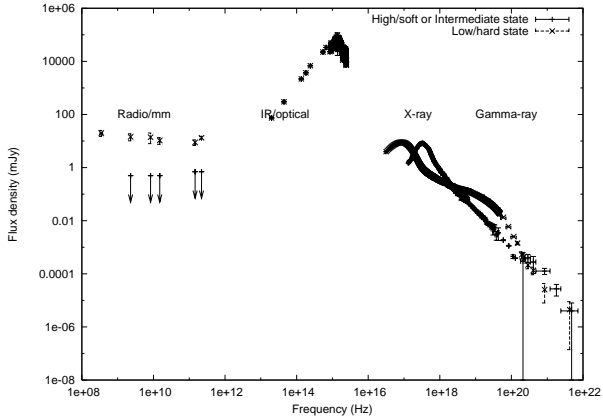,angle=270,width=8cm}
\caption{The change in the broadband spectrum of the black hole X-ray
  binary Cygnus X-1 between two X-ray 'states'. In the 'low/hard'
  state the accretion disc has a lower temperature, with a strong
  'excess' around 100 keV, and the radio emission is 'on', with a
  relatively flat spectrum. This probably corresponds to the
  generation of a self-absorbed $\sim$conical jet during phases when
  we do not see a bright accretion disc extending to the innermost
  stable circular orbit. In the softer state the X-ray spectrum
  is dominated by a hotter accretion disc, probably closer to the
  black hole, and the radio emission ($\rightarrow$ jet) is 'quenched'.
}
\end{figure}

\subsubsection{Neutron star X-ray binaries}

The study of the disc-jet coupling in neutron star (NS) X-ray binaries
has lagged behind that of the black holes in recent years. The reason
for this is that the NS X-ray binaries are considerably less 'radio
loud' than the black holes (Fender \& Kuulkers 2001; Migliari et
al. 2003 and in prep). Even when considering the mass term in the 'Fundamental
plane' (see below), they appear to be less efficient at producing
radio emission for a given X-ray flux. 

Despite complex patterns of behaviour in X-rays on relatively short
(hours) timescales, the radio:X-ray plane for NS X-ray binaries
appears similar to Fig 3, but with the neutron stars occupying a
region a factor of $\sim 30$ below the black holes (Migliari et al. in
prep). This strongly implies that the study of these NS jet sources
can tell us something about the black holes, since

\begin{itemize}
\item{The apparent similarity in the patterns tells us that the global
properties of the disc-jet coupling in black holes are as a result of
the accretion flow and not something unique to black holes themselves}
\item{The clear difference in radio power indicates that some
  difference between NS and black hole X-ray binaries (surface,
  radiation, magnetic field) is enough to affect the observed jet
  power}
\end{itemize}

Furthermore, we already have evidence that the most relativistic flows
from binaries are in fact from the neutron star systems (see
below). The detailed study of jets from accreting neutron stars is
likely, for the reasons given above, to be crucial in our
understanding of the jet formation process in black holes.

\begin{figure*}
\psfig{figure=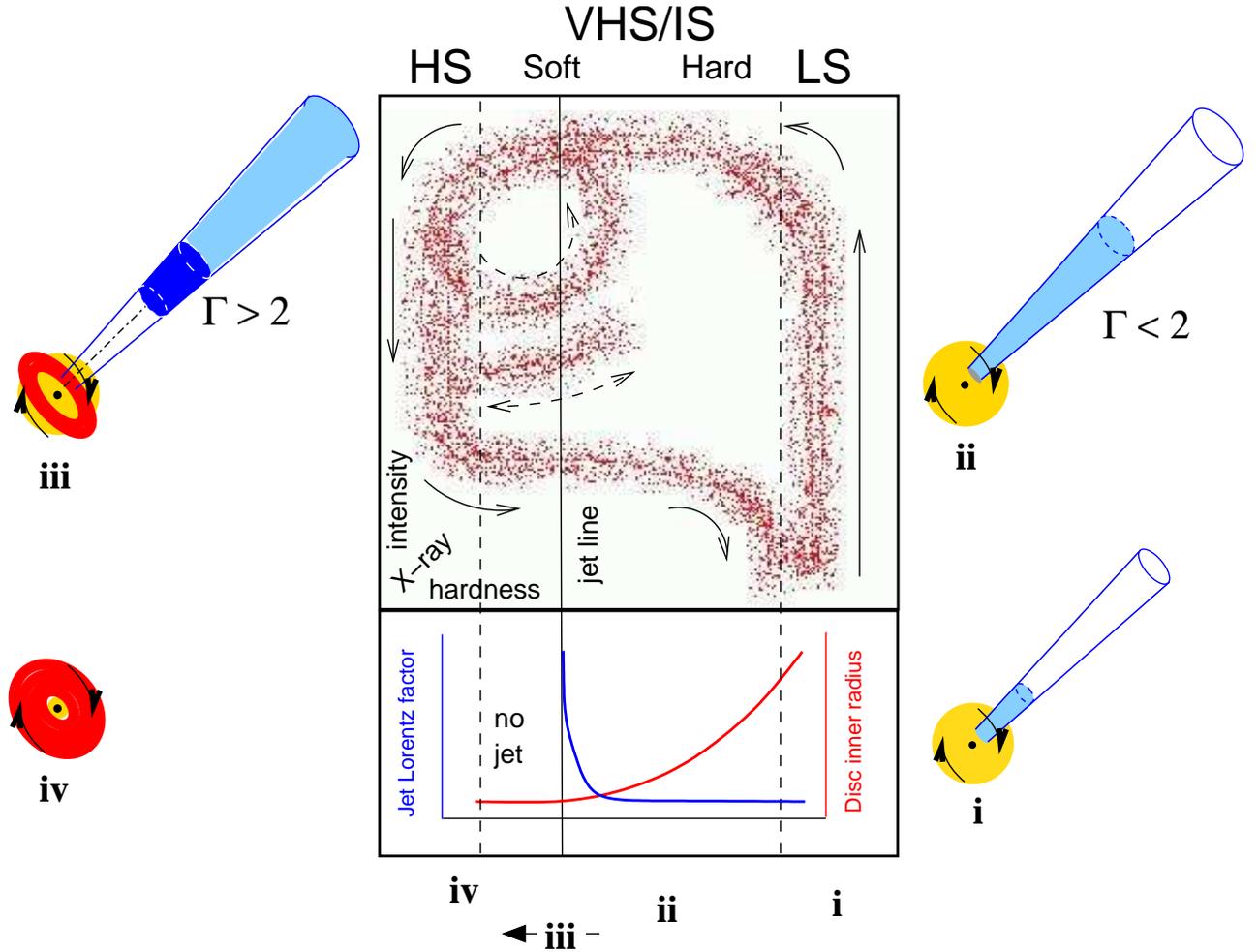,width=17cm}
\caption{A schematic of our simplified model for the jet-disc coupling
  in black hole binaries. The central box panel represents an X-ray
  hardness-intensity diagram (HID); 'HS' indicates the `high/soft
  state', 'VHS/IS' indicates the 'very high/intermediate state' and
  'LS' the 'low/hard state'. In this diagram, X-ray hardness increases
  to the right and intensity upwards. The lower panel indicates the
  variation of the bulk Lorentz factor of the outflow with hardness --
  in the LS and hard-VHS/IS the jet is steady with an almost constant
  bulk Lorentz factor $\Gamma < 2$, progressing from state {\bf i} to
  state {\bf ii} as the luminosity increases. At some point -- usually
  corresponding to the peak of the VHS/IS -- $\Gamma$ increases
  rapidly producing an internal shock in the outflow ({\bf iii})
  followed in general by cessation of jet production in a
  disc-dominated HS ({\bf iv}). At this stage fading optically thin
  radio emission is only associated with a jet/shock which is now
  physically decoupled from the central engine.  As a result the solid
  arrows indicate the track of a simple X-ray transient outburst with
  a single optically thin jet production episode. The dashed loop and
  dotted track indicate the paths that GRS 1915+105 and some other
  transients take in repeatedly hardening and then crossing zone {\bf
  iii} -- the 'jet line' -- from left to right, producing further
  optically thin r adio outbursts. Sketches around the outside
  illustrate our concept of the relative contributions of jet (blue),
  'corona' (yellow) and accretion disc (red) at these different
  stages.}
\label{fbg}
\end{figure*}

\subsection{X-ray binaries as tools to understand AGN}

Much has been made in the past decade of the apparent analogy between
the relativistic jets produced by supermassive ($10^6 M_{\odot} \leq M_{\rm BH}
\leq 10^{10} M_{\odot}$) black holes (AGN) and those produced in BH
XRBs ($3 M_{\odot} \leq M_{\rm BH} \leq 20 M_{\odot}$) -- hence the
popular name `microquasars' for this latter class of object.  However,
for most of this period the comparisons remained largely
phenomenological. However, in the past year all this has changed as
{\em quantitative} scalings between jets, and between X-ray and radio
power in black holes of {\em all masses} and {\em all accretion rates}
have emerged (Heinz \& Sunyaev 2003; Merloni, Heinz \& di Matteo 2003;
Falcke, K\"ording and Markoff 2004).  Specifically, Merloni, Heinz \&
di Matteo (2003) and Falcke. K\"ording \& Markoff (2004) have
demonstrated the existence of a 'fundamental plane' of black hole
activity in the 3D parameter space of black hole mass ($M_{\rm BH}$),
X-ray luminosity ($L_{\rm X}$) and radio ($L_{\rm radio}$)
luminosity (Fig 6). This fundamental plane spans a range in $>10^8$ in $M$,
albeit currently with a lot of scatter. Maccarone, Gallo \&
Fender (2003) have recently rebinned the data set of Merloni, Heinz \& di
Matteo to demonstrate the 'quenching' of jets in the same fractional
Eddington range as known to occur in XRBs, suggesting the disc-jet
phenomenology may also be the same.

Detailed quantitative comparisons are only just beginning to be made;
and will no doubt be the subject of many future research papers. At
the very roughest level, it is tempting to associate the (disputed)
`radio loud' and `radio quiet' dichotomy observed in AGN with
jet-producing (hard and transient) and non-jet-producing (soft) states
in X-ray binaries (see e.g. Maccarone, Gallo \& Fender 2003),
including of course the effect of the mass term.  Furthermore perhaps
FRI jet sources can be associated with the low/hard state and FRIIs
with transients. Meier (1999; 2001) has considered jet production
mechanisms in both classes of object, and drawn interesting
parallels. Gallo, Fender \& Pooley (2003; amongst others!) have made a
qualitative comparison between FRIs and low/hard state black hole
X-ray binaries and FRIIs and transients.

It is interesting to note that the short timescale disc-jet coupling
observed in GRS 1915+105 (Pooley \& Fender 1997; Eikenberry et
al. 1998; Mirabel et al. 1998; Klein-Wolt et al. 2001), in its most
basic sense -- that radio events are preceded by a `dip' and
associated spectral hardening in the X-ray light curve -- may also
have an analog in AGN: Marscher et al. (2002) have reported
qualitatively similar behaviour in 3C 120.

\begin{figure*}
\psfig{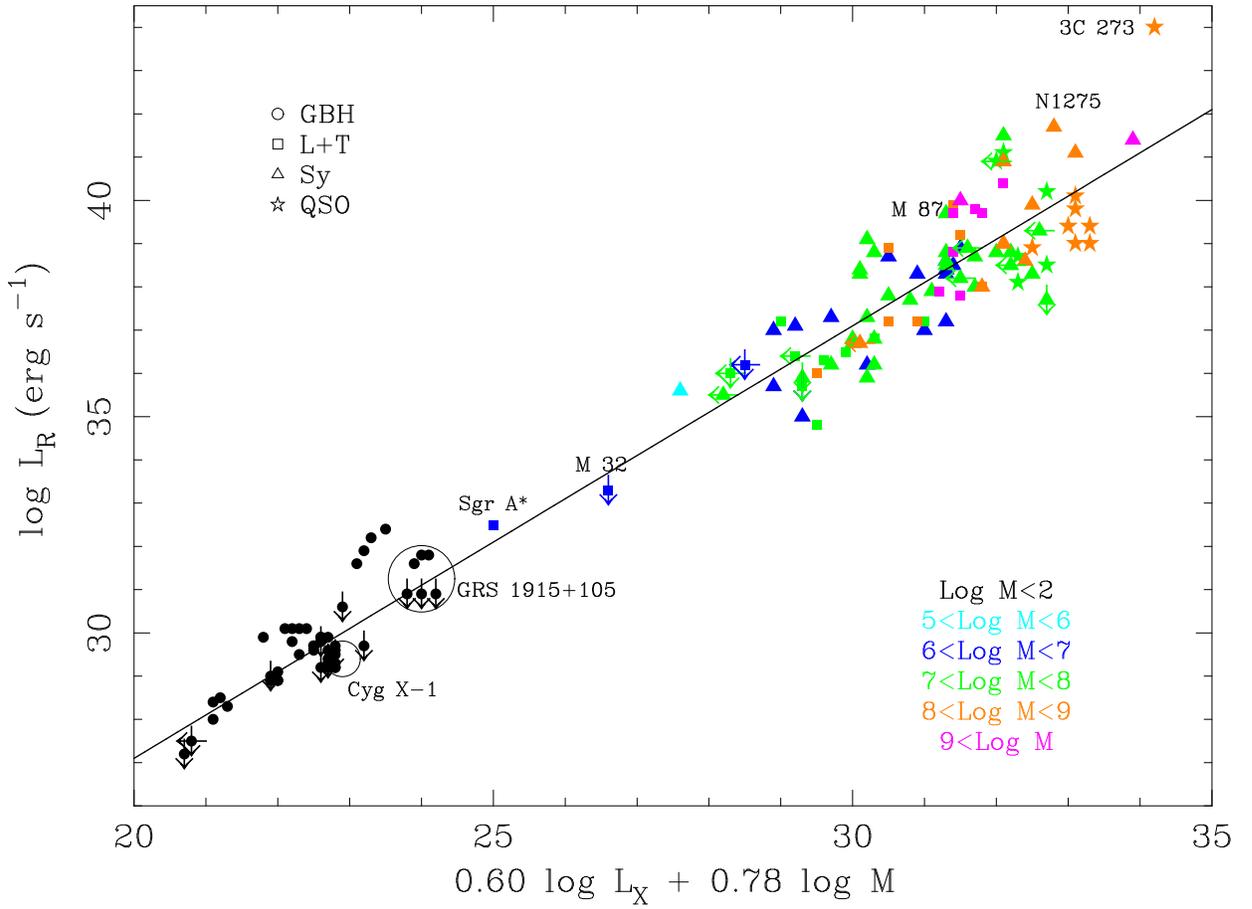}
\caption{The fundamental plane of black hole activity (Merloni, Heinz
  \& di Matteo 2003; see also Falcke, K\"ording \& Markoff
  2004). Combined data from both X-ray binaries and AGN indicates the
  existence of an approximate plane in the $L_{\rm radio}$:$L_{\rm
  X}$:$M_{\rm BH}$ parameter space. This underlines the
  scale-invariance of the accretion:outflow coupling (Heinz \& Sunyaev
  2003) and demonstrates that results from X-ray binaries and AGN may
  be compared {\em quantitatively}.
}
\end{figure*}

\section{Studying X-ray binaries with SKA}

The potential offered by the SKA to study the formation of jets in
X-ray binaries, and their relation to the accretion process (by means
of simultaneous X-ray observations) is immense. With a two orders of
magnitude sensitivity leap and VLBI-scale angular resolution, we
expect to regularly resolve relativistic events from a large number of
X-ray binary systems, as well as detecting unresolved radio emission
from radio cores at very low luminosities.

\subsection{Monitoring bright sources}

A handful of X-ray binaries remain semi-continuously in very bright
and variable X-ray and/or radio states ('outburst'). These systems
include SS 433, Cygnus X-3, GRS 1915+105 (see Fig three inset
(iii))and Circinus X-1 (see Fig 8). For such sources, with typical
flux densities 1---1000 mJy at GHz wavelengths, and previously
resolved jet-like structures, we may expect low-duty-cycle radio
monitoring to result in movies capturing the jet formation and
propagation in exquisite detail.

Furthermore, bright (close to Eddington at peak) new X-ray transients
are expected to produce (sequences of) strong radio outbursts
(probably corresponding to internal shocks; see Fig {\ref{fbg}}) with
a frequency typically around $\sim 1$/year. Such sources when resolved
typically reveal ejections with projected velocities in the range
1--15$c$, decelerating at larger distances from the core (see
e.g. Fender 2004 and references therein; Fender, Belloni \& Gallo
2004; Kaaret et al. 2003). These jets typically fade rather rapidly
(with e-folding times of hours to weeks) and become unobservable due
to the sensitivity of the radio array before we have observed them
doing anything 'interesting' such as interacting with the ISM. SKA has
the potential to track these sources much further from the core,
observing such interactions and decelerations. Furthermore, the short
integration time required to make 'snapshot' monitoring observations
of recent transients will undoubtedly facilitate the discovery of more
'rebrightening' events such as those observed in the case of XTE
J1550-564 years after the initial outburst (Corbel et al. 2002).

The jets from these sources are also known to display both linear and
circular polarisation (e.g. Fender et al. 1999,2000a,b; Macquart \&
Fender 2004). Linear polarisation observations provide unique
information on the degree or ordering and orientation of the magnetic
field within the ejecta. The situation is clearly not straightforward
-- spatially resolved linear polarisation images of GRS 1915+105 made
with MERLIN (Fender et al. 1999) revealed a field which was different
each day. Other, unresolved, events from the same source observed with
ATCA have revealed both constant polarisation angles and 'rotator'
events (Fender et al. 2002a,b -- see inset (iii) in Fig 3). Circular
polarisation (Macquart \& Fender 2004) offers a complex but
potentially very powerful insight into the composition of relativistic
jets. For example, the detection of a circular polarisation spectrum
of the form $V/I \propto \nu^{-0.5}$ from a homogenous, optically
thin, synchrotron source would be strong evidence for a baryonic
component in the jet plasma, something which (with the exception of SS
433) has eluded observations so far.

\subsection{Resolving 'steady' jets in hard X-ray states}

As outlined above (see e.g. Fender 2004 for a more detailed
discussion), black holes in 'hard' X-ray states seem to be
ubiquitously associated with steady jet production (see insets (i) and
(ii) in Fig 3). The power and physical conditions in these jets are
central to our understanding of how accretion proceeds near a black
hole. The overwhelming majority of black holes of all masses are
likely to exist in this state.

Observations of synchrotron time lags (e.g. Mirabel et al. 1998;
Fender et al. 2002a) as well as direct imaging (Fuchs et al. 2003)
indicate that the compact jet in GRS 1915+105 has a physical size of
around $10^{14-15}$ cm at GHz wavelengths for a flux density in the
range 40--120 mJy. If we assume that this physical size corresponds to
the distance from the black hole to the $\tau \sim 1$ surface (where
$\tau$ is the optical depth) in a self-absorbed jet (e.g. Blandford \&
Konigl 1979; see also chapter by Falcke, K\"ording \& Nagar), and that
all compact jets have approximately the same brightness temperature
(as seems to be the case for AGN -- e.g. Ghisellini et al. 1993), then
we can estimate the physical size of such a jet as

\[
r_{\rm GHz} \sim 5\times10^{14} \left(\frac{S_{\nu}}{40{\phantom{0}}{\rm
    mJy}}\right)\left(\frac{d}{11\phantom{0}{\rm kpc}}\right)^2
    \phantom{00}{\rm cm}
\]

\begin{figure}
\psfig{figure=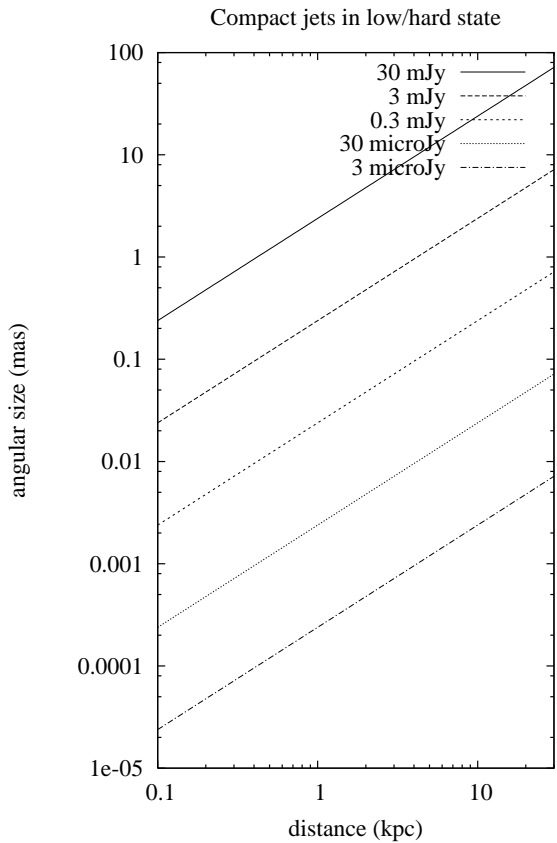,width=8cm,angle=0}
\caption{Estimates (GHz) size of compact 'core' jets from black holes for
  GHz radio flux densities of 30, 3 and 0.3 mJy. A typical hard state
  has a flux density of $\geq$ few mJy at a distance of 2--20 kpc and
  so is potentially resolveable with high sensitivity VLBI. 
}
\label{cores}
\end{figure}

This corresponds to an angular size of

\[
\alpha \sim 35 \left(\frac{S_{\nu}}{40{\phantom{0}}{\rm
    mJy}}\right)\left(\frac{d}{11\phantom{0}{\rm kpc}}\right) 
    \phantom{00}{\rm mas}
\]

\noindent which means that for a given radio flux density, a distant
source will be more easy to resolve than a nearby one, because it is
more powerful. This function is plotted in Fig {\ref{cores}}. A
typical hard state X-ray transient will have a flux density of $\geq$
a few mJy and lie at a distance of 2--20 kpc. Such sources are
potentially resolveable with high sensitivity VLBI. Note that Gallo,
Fender \& Hynes (2004) have recently measured the radio spectrum of
the black hole X-ray binary V404 Cyg in 'quiescence' and have found it
to be similar to that of Cyg X-1, supporting the interpretation of
low-level radio emission originating in a compact jet.

The handful of measurements of linear polarisation from hard state
sources indicate a typical level of $\sim 2$\% (Fender
2001). Furthermore, the polarisation angle seems to be aligned with
the jet axis, for at least one source (Gallo et al. 2004). Careful
monitoring of variations with time of the linear polarisation vector
could, potentially, reveal the precession of steady jets whose angular
extent will never allow them to be directly resolved (the effect 
can be quite large: in SS 433 the jets precess with a
half-angle of $\sim 20^{\circ}$).

\subsection{Black holes in 'quiescence' -- probing advection}

The power-law correlation found for low/hard state black hole binaries
(Corbel et al. 2003; Gallo et al. 2003), combined with X-ray
measurements, allows us to predict the radio luminosities of
'quiescent' black hole systems (Gallo et al. 2003). These systems are
believed to be strongest examples of 'advection dominated accretion
flows' (ADAFs) and typically have soft X-ray luminosities around
$L_{\rm X} \sim 10^{-8}$ Eddington. The predicted quiescent
radio flux densities are in the range 1--30 $\mu$Jy (Gallo et
al. 2003). 

Fender, Gallo \& Jonker (2003) demonstrated that the same power-law
relation, if maintained to such low luminosities, would be very strong
evidence for the existence of 'jet dominated states' in which the
dominant power output channel for the liberated accretion energy is a
radiatively-inefficient jet, and not X-ray emission. Measurements of
the broadband spectra of these 'quiescent' black holes in the way done
for Cygnus X-1 at higher luminosity (Fig 4) is strongly limited by our
radio sensitivity (and also $\gamma$-ray, but this component is rather
unexplained anyhow!). We have recently obtained a four-frequency radio
spectrum of a black hole X-ray binary, V404 Cyg, with a mean X-ray
luminosity of $\sim 10^{-6}$ Eddington (Gallo, Fender \& Hynes
2004). The radio spectrum is flat, supporting a jet interpretation,
and the mean radio flux density is $\sim 0.3$ mJy.

The SKA will allow us to measure the broadband radio spectra of the
truly 'quiescent' sources, which are expected to be one to two orders
of magnitude fainter than V404 Cyg (Gallo, Fender \& Pooley 2003).
which may be combined with IR/optical/X-ray
measurements to allow us to estimate the jet power at such low
accretion rates. Furthermore, a galactic plane survey with SKA may
well discover a population of unknown quiescent accreting objects
which stand out due to their compact size and flat spectra, but would
not stand out in X-ray surveys. Inspection of Fig {\ref{cores}} shows
that its rather unlikely we will be able to directly resolve the jets
from such weak sources, with anticipated micro-arcsecond angular
sizes.

\subsection{Intermediate mass black holes in globular clusters?}

It has been argued that globular clusters are likely to contain in
their centres 'intermediate mass' black holes, formed via stellar
mergers. The masses of these central black holes may be as high as
$10^{-3}$ of the cluster mass. 

Maccarone (2004) has recently demonstrated that the fundamental plane
of black hole activity implies that radio emission is a much more
stringent test for the existence of these intermediate mass black
holes in the centres of globular clusters than X-ray observations.
For accretion from the ambient medium in these clusters, radio flux
densities in the range 1--100 $\mu$Jy are expected. The SKA will be
able to achieve such sensitivities with relative ease, and strong
tests for the existence of the postulated central black holes in
globular clusters may be undertaken. 

\subsection{Extremes and unexpected phenomena}

Just as we are preparing to settle into a pattern of observation
designed to confirm and consolidate our ideas, along come, thankfully,
some unexpected phenomena, There is no doubt that SKA will discover
more such phenomena, but the examples outlined below both illustrate
the surprises and present areas in which SKA will be able to make
significant progress.

\subsubsection{Ultrarelativistic flows from neutron stars}

\begin{figure*}
\centerline{\psfig{figure=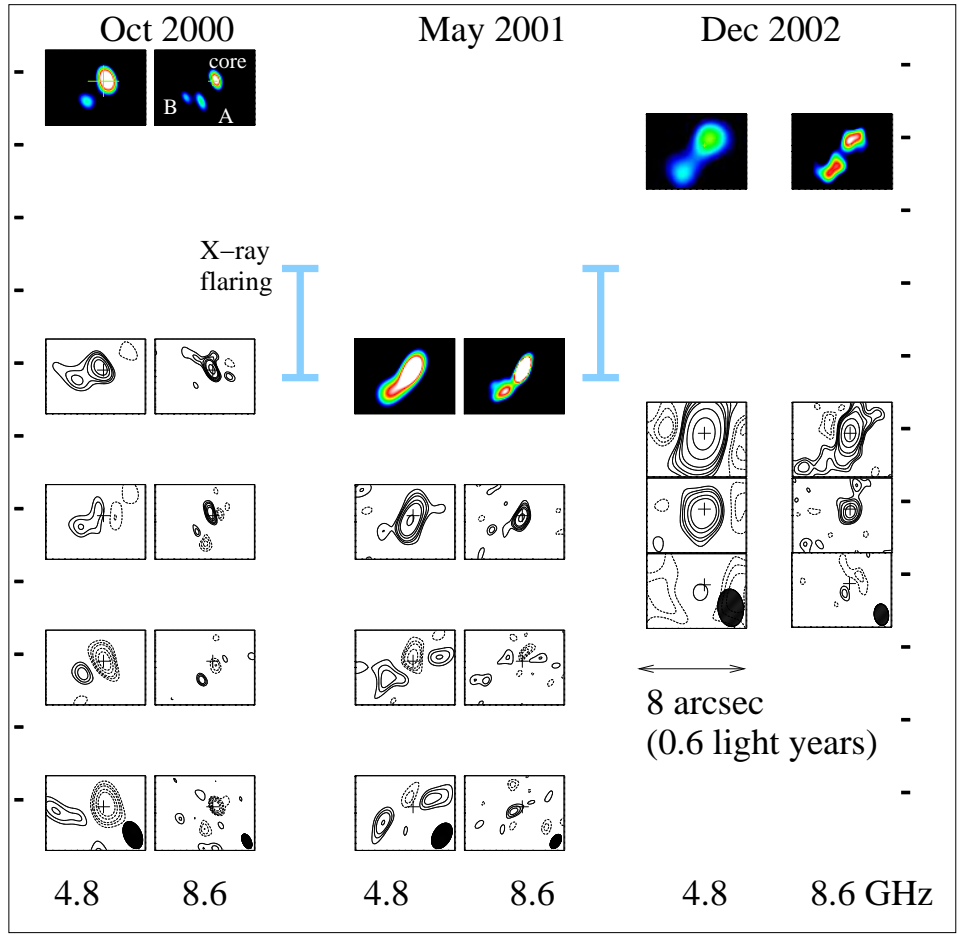,width=14cm,angle=270}}
\caption{An ultrarelativistic outflow: sequences of radio observations
of Circinus X-1 in October 2000, May 2001 and December 2002. At each
epoch, observations were made simultaneously at 4.8 and 8.6 GHz. White
tickmarks indicate time steps of one day; the blue bar indicates the
time of the X-ray flaring as observed by the {\em Rossi} X-ray Timing
Explorer All-Sky Monitor. In October 2000 and May 2001 the
observations were spaced every two days; in December 2002 they are
daily. At each epoch the {\em u-v} coverage of the radio observations
is identical for each image; maps in October 2000 and May 2001 are
`uniformly weighted', those in December 2002 are `naturally
weighted'. The crosses indicate the location of the binary `core' from
\cite{fen98}, and their size is proportional to the `core' radio flux
density.  The observations reveal that following the X-ray flaring the
extended radio structure brightens on timescales of days. The apparent
velocities associated with this expansion are $\geq 15c$, indicating
an underlying ultrarelativistic flow.  }
\label{cirx1}
\end{figure*}

Fomalont et al. (2001a,b) and Fender et al. (2004; see Fig
{\ref{cirx1}}) have reported evidence for the existence of highly
relativistic, but essentially 'unseen' outflows from neutron stars
accreting at high rates. In the case of Sco X-1 (Fomalont et
al. 2001a,b) the outflow is observed to move with a bulk Lorentz
factor $\Gamma \geq 3$; in the case of Cir X-1 (Fender et al. 2004)
the lower limit is much greater $\Gamma \geq 15$. This is probably due
to the small angle of the Cir X-1 jet to the line of sight and
suggests that at least all six neutron star 'Z' sources (of which Sco
X-1 is the archetype) are producing such 'ultrarelativistic' flows. SS
433 may also be producing such a flow, resulting in transient
energisation of the well-known knots which move at $\sim 0.26c$
(Migliari et al. 2004).

These flows are rather unexpected in the simple, widely-accepted
framework in which jets reflect the escape velocity in the region in
which they are formed, since this should not be more than $\sim 0.4c$
for a neutron star (e.g. Livio 1999). The nature of these flows also
remains a mystery; for example they may be bright but so fast they
they're always beamed out of our line of sight, or they may be 'cold'
or highly radiatively inefficient and only ever observed via their
interactions with other components. Their importance lies in the fact
that they clearly demonstrate that whatever it is that is necessary
for the formation of highly relativistic flows, it is not something
which is unique to black holes. The very rapid variability of the jet
in Cir X-1, with a projected velocity $\geq 15c$ and hints in the
radio maps of a very transient structure (Fender et al. 2004)
highlight the need for high sensitivity snapshots such as will be
provided by SKA. 

\subsubsection{Jets from obscured X-ray binaries as $\gamma$-ray sources}

Paredes et al. (2000) reported the discovery of a powerful, seemingly
persistent, radio jet from the massive X-ray binary LS 5039. This
system had however a strange broadband spectrum, being weak in X-rays
but potentially associated with an unidentified EGRET gamma-ray
source.

Furthermore, INTEGRAL observations have revealed what might be a
population of X-ray binaries undergoing such intense local absorption
(by gas, and maybe also dust, local to the binary system), that they
have remained undetected by previous soft X-ray surveys. In such
sources, whatever the source of the X-ray absorption, the jet is
likely to extend beyond this region, and they will remain detectable
as radio sources. A deep, multi-frequency radio survey of the galactic
plane with SKA may well discover many such sources.

\subsection{Beyond the Milky Way}

\begin{figure}
\psfig{figure=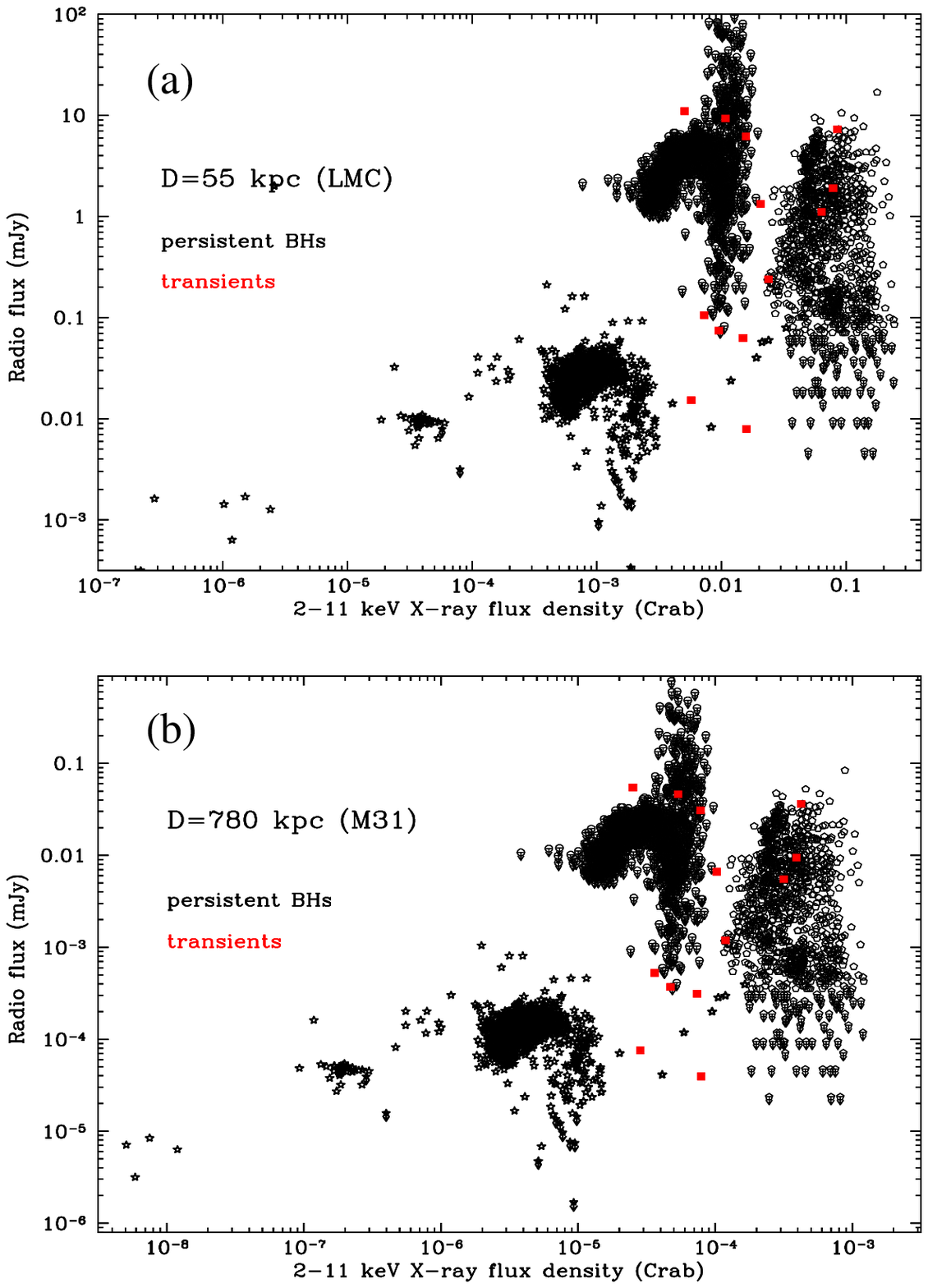,width=8cm,angle=0}
\caption{Top panel (a): As Fig {\ref{gfp}}, but scaled for a distance
    of 55 kpc, corresponding to the LMC, and also approximately
    appropriate for the SMC and Sagittarius, Sculptor, Sextans, Draco
    and Ursa Minor dwarf spheroidals. Clearly transient radio sources
    in these nearby galxies will be easily detectable with SKA, as
    will the brightest of the 'steady' (hard X-ray state)
    sources. Lower panel (b): as (a) but scaled to M31 / M33. Even at this
  distance, corresponding effectively to the radius of the Local Group
  of ($\sim 40$) galaxies, bright radio transients should be readily
  detectable with SKA. The field of M31 / M33 in particular, as it will
  no doubt be regularly monitored in X-rays, will provide fertile
  ground for comparing the disc-jet coupling with that in our own galaxy.
}
\label{gallo2}
\end{figure}

No X-ray binary system has clearly been unambiguously detected in the
radio band beyond the Milky Way, despite the observations of hundreds
of such sources in external galaxies (e.g. Fabbiano \& White 2004).

The radio:X-ray plane for galactic black hole binaries presented in
Fig 3 may be simply scaled to the distances of local galaxies to see
how SKA may help our cause in this area. The top panel of Fig
{\ref{gallo2}} shows the plane scaled to the distance of the LMC (and,
effectively, the SMC). Bright transients in these two satellite
galaxies are already potentially visible with e.g. ATCA, and would be
easily accessible to SKA. Perhaps more importantly, so would the
steady 'low/hard' state sources, and the brightest neutron
stars. Since we know of at least one of each class of object in the
LMC (LMC X-3 and X-2 respectively), we could test the possible
dependence of the disc-jet in both black holes and neutron stars as a
function of metallicity.

Extending our view across the Local Group of galaxies to M31 (Fig
{\ref{gallo2}}, lower panel), we can see that relatively little effort
still is required to detect bright transients at this distance. Even
the brightest steady sources would be detectable with sufficiently
long ($\sim 1$ day) integrations. This is a very exciting prospect,
allowing us to calibrate the radio and X-ray correlation without the
strong distance uncertainties which plague us within the Milky Way.

\begin{figure}
\psfig{figure=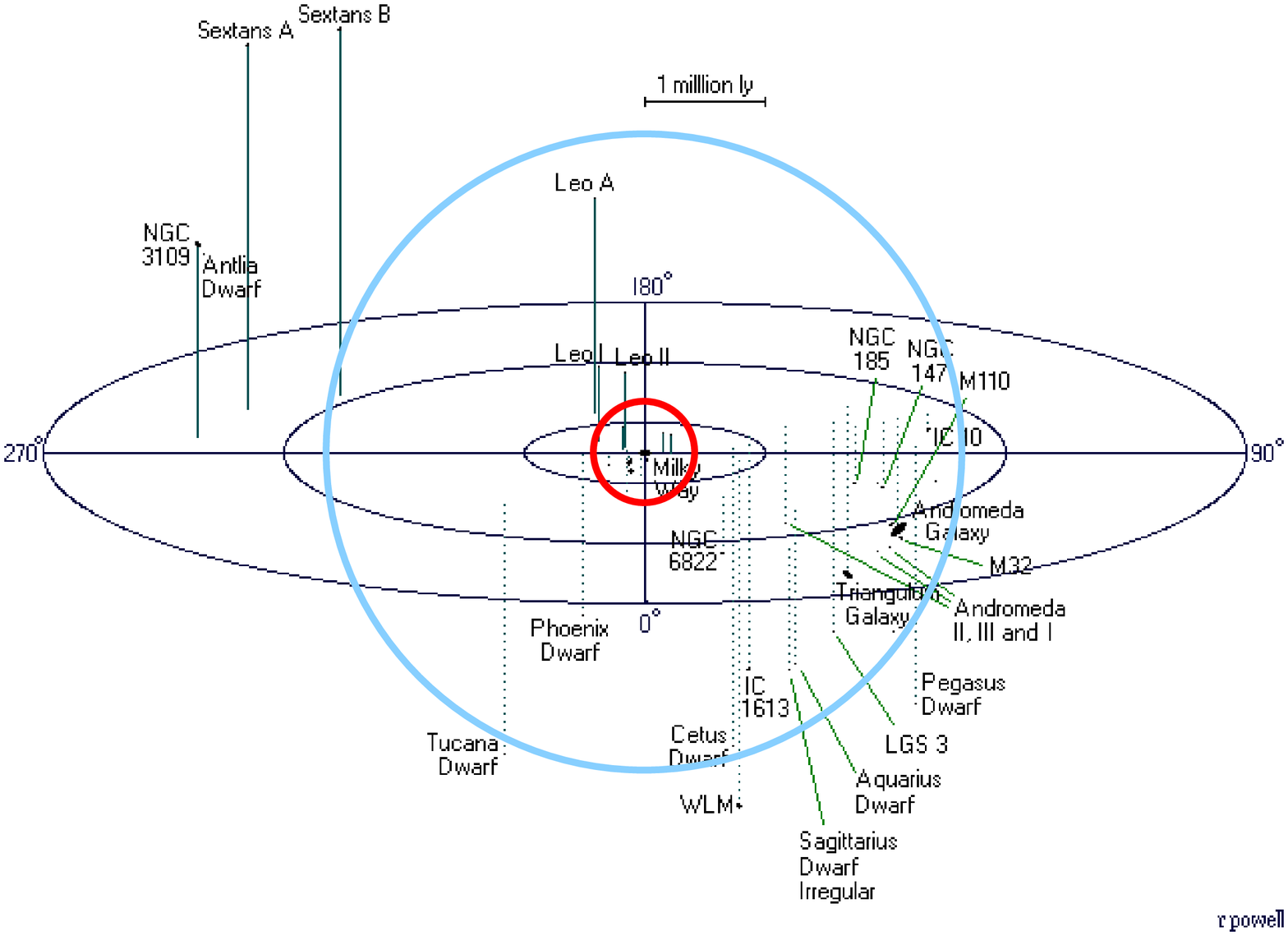,width=8cm,angle=0}
\caption{The local group of galaxies. The SKA will detect persistent
  (hard X-ray state) binaries in the nearest galaxies (LMC, SMC and
  other nearby dwarf galaxies). Transient sources will be detectable
  throughout the local group, covering approximately 40 galaxies and
  $\sim 10^{11}$ stars.}
\label{lg}
\end{figure}

Fig {\ref{lg}} illustrates in summary the 'reach' of SKA for the study of X-ray
binary systems. The inner dark circle indicates those regions for
which the detection of radio emission from both transients and steady
sources would be more or less trivial. For sources at the distance of
the lighter, larger, ring (encompassing nearly all the mass of the
Local Group), transients would still be relatively easy to detect, and
steady sources would be accessible via long integrations.

\section{Requirements for the SKA}

The exact specifications for the SKA are slowly being assembled. Such
a large step in sensitivity will have enormous benefits for this field
of research, whatever the precise design. Nevertheless, it is useful
to lay out what we have anticipated for the design, and what will be
important for the study of jets from X-ray binaries.

\begin{itemize}
\item The discussions in this document have been based upon a SKA which has,
a sensitivity at GHz frequencies which is about two orders of
magnitude better than the current VLA / ATCA / WSRT. This corresponds
to, for example, a 10-min sensitivity at 5 GHz of $\sim 1 \mu$Jy.

\item The angular resolution of the SKA required for the various studies
outlined in this paper should be comparable to that obtained currently
at a few GHz with VLBI. This is not only necessary for the direct
imaging of jets for sources within our own galaxy but also for
clearly resolving individual sources from the background at larger
distances.

\item The frequency range of the proposed SKA is also important for the
context of X-ray binary studies. A broad frequency range will allow
the spectrum of the radio emission, which we have found to be
dependent on the X-ray state and -- possibly -- to the X-ray
luminosity within that state, to be well determined. 

\item Simultaneous multi-frequency capability will be important.
The
observations of time delays in the synchrotron emission from the
powerful jet source GRS 1915+105 (e.g. Mirabel et al. 1998; Fender et
al. 2002a) give us a direct insight into the scale of the jet
unobtainable by other means.

\item Good polaristion sensitivity ('purity'). Both linear and
  circular polarisation provide important clues about the structure
  and composition of the synchrotron emitting components which cannot
  be otained in any other way. 
\end{itemize}

\section*{Acknowledgements}

The author would like to thank Elena Gallo for producing Fig
{\ref{gallo2}}. Robert Braun, Heino Falcke, Ben Stappers and many
others have contributed indirectly to this work via many useful
discussions about the detection of transient radio sources with the
next generation of radio telescopes.

\end{document}